\documentclass[9pt,twocolumn,twoside]{pnas-new}

\templatetype{pnasinvited} 

\newenvironment{myepigraph}
  {\par\hfill\itshape
   \begin{tabular}{@{}r@{\hspace{2em}}}} 
  {\end{tabular}\par\medskip}

\begin{document}

\title{Automating the Practice of Science – Opportunities, Challenges, and Implications}

\author[a]{Sebastian Musslick*}
\author[b]{Laura K. Bartlett}
\author[c]{Suyog H. Chandramouli}
\author[d]{Marina Dubova}
\author[b,e]{Fernand Gobet}
\author[f]{Thomas L. Griffiths}
\author[g]{Jessica Hullman}
\author[h]{Ross D. King}
\author[i]{J. Nathan Kutz}
\author[j]{Christopher G. Lucas}
\author[k]{Suhas Mahesh}
\author[l]{Franco Pestilli}
\author[m]{Sabina J. Sloman}
\author[d]{William R. Holmes}

\affil[a]{Institute of Cognitive Science, Osnabrück University, 49090 Osnabrück, Germany; Department of Cognitive, Linguistic, \& Psychological Sciences, Brown University, Providence, RI 02912, USA, \href{https://orcid.org/0000-0002-8896-639X}{ORCID}}
\affil[b]{Centre for Philosophy of Natural and Social Science, London School of Economics, Lakatos Building, Houghton Street, London, WC2A 2AE, UK, \href{https://orcid.org/0000-0001-5202-4504}{ORCID}}
\affil[c]{Department of Information and Communications Engineering, Aalto University, P.O. Box 11000 (Otakaari 1B) FI-00076 AALTO, Finland; Department of Computing Science, University of Alberta, 8900 114 St NW, Edmonton, AB T6G 2S4, Canada \href{https://orcid.org/0000-0002-3450-1909}{ORCID}}
\affil[d]{Cognitive Science Program, Indiana University, 1101 E 10th St, Bloomington, IN 47405, USA, \href{https://orcid.org/0000-0001-5264-0489}{ORCID}}
\affil[e]{School of Psychology, University of Roehampton, London, SW15 4JD, UK \href{https://orcid.org/0000-0002-9317-6886}{ORCID}}
\affil[f]{Departments of Psychology and Computer Science, Princeton University, Princeton, NJ, USA, \href{https://orcid.org/0000-0002-5138-7255}{ORCID}}
\affil[g]{Department of Computer Science, Northwestern University, IL, USA, \href{https://orcid.org/0000-0001-6826-3550}{ORCID}}
\affil[h]{Department of Chemical Engineering and Biotechnology, University of Cambridge, Cambridge, CB3 0AS, UK; Department of Computer Science and Engineering, Chalmers University of Technology, Gothenburg, 412 96, Sweden; \href{https://orcid.org/0000-0001-7208-4387}{ORCID}}
\affil[i]{Department of Applied Mathematics and Electrical and Computer Engineering, University of Washington, Seattle USA 98195, \href{https://orcid.org/0000-0002-6004-2275}{ORCID}}
\affil[j]{School of Informatics, University of Edinburgh, 10 Crichton St., EH8 9AB, United Kingdom, \href{https://orcid.org/0000-0002-6655-8627}{ORCID}}
\affil[k]{Department of Materials Science and Engineering, University of Toronto, Canada, \href{https://orcid.org/0000-0002-3897-7963}{ORCID}}
\affil[l]{Department of Psychology and Department of Neuroscience, The University of Texas, Austin, TX, USA, \href{https://orcid.org/0000-0001-5264-0489}{ORCID}}
\affil[m]{Department of Computer Science, University of Manchester, M13 9PL, UK \href{https://orcid.org/0000-0002-2469-049}{ORCID}}
\leadauthor{Musslick}

\authorcontributions{Author contributions.
Based on the perspectives submitted by all authors, SM and WRM outlined the initial draft, and SM wrote the initial draft. All authors revised the initial draft and contributed to revisions that were incorporated into the final version.}
\correspondingauthor{*To whom correspondence should be addressed. Sebastian Musslick, Institute of Cognitive Science, Wachsbleiche 27, 49090 Osnabrück, Germany, sebastian.musslick$@$uos.de}


\begin{abstract}
Automation transformed various aspects of our human civilization, revolutionizing industries and streamlining processes. In the domain of scientific inquiry, automated approaches emerged as powerful tools, holding promise for accelerating discovery, enhancing reproducibility, and overcoming the traditional impediments to scientific progress. This article evaluates the scope of automation within scientific practice and assesses recent approaches. Furthermore, it discusses different perspectives to the following questions: Where do the greatest opportunities lie for automation in scientific practice?; What are the current bottlenecks of automating scientific practice?; and What are significant ethical and practical consequences of automating scientific practice? By discussing the motivations behind automated science, analyzing the hurdles encountered, and examining its implications, this article invites researchers, policymakers, and stakeholders to navigate the rapidly evolving frontier of automated scientific practice.
\end{abstract}

\dates{This manuscript was compiled on \today}
\doi{\url{www.pnas.org/cgi/doi/10.1073/pnas.XXXXXXXXXX}}

\maketitle
\thispagestyle{firststyle}
\ifthenelse{\boolean{shortarticle}}{\ifthenelse{\boolean{singlecolumn}}{\abscontentformatted}{\abscontent}}{}

\begin{myepigraph}
``Though the world does not change with a change of paradigm,\\ the scientist afterward works in a different world.''\\ – Thomas S. Kuhn, The Structure of Scientific Revolutions
\end{myepigraph}

\noindent Automation is transforming every domain of scientific inquiry, from the study of functional genomics in biology \cite{king_functional_2004,king_automation_2009} to the derivation of conjectures in mathematics \cite{raayoni_generating_2021,davies_advancing_2021}. Recent advances in automation are accelerating hypothesis generation in chemistry \cite{de_almeida_synthetic_2019,jumper_highly_2021,lindsay_applications_1980,boiko_autonomous_2023}, material discovery in materials science \cite{merchant_scaling_2023,szymanski_autonomous_2023}, and theory development in psychology \cite{peterson_using_2021}. These breakthroughs are not only garnering attention but also an uptick in funding and prizes dedicated to the automation of scientific practice \cite{us_department_of_energy_scientific_2023,velasquez_foundation_2023,kitano_nobel_2021}. Furthermore, concurrent advancements in artificial intelligence, software, and computing hardware are setting the stage for even more extensive automation within the scientific process \cite{zenil_future_2023,birhane_science_2023,wang_scientific_2023}.

The impact of automation in industry serves as a parallel to its potential in science. In the early 20th century, industrial automation began with mechanized assembly lines, revolutionizing manufacturing efficiency and output. The introduction of robotics and computer-aided manufacturing marked another leap, enabling precision and consistency previously unattainable by human labor. Today, industry-wide automation facilitates not just cost-efficient mass production, but also customized, adaptable, and intelligent manufacturing processes. This evolution demonstrates the capacity of automation to radically redefine operational paradigms.

Drawing parallels to scientific practice, one can anticipate a similar trajectory of profound change, where automation could accelerate discovery, reshape research methodologies, and redefine the very nature of scientific inquiry. At the same time, automation in industry had significant impacts on workers and the kind of products that dominate the marketplace. It is thus important to consider parallel impacts in the scientific setting which may have negative consequences for science and society. 

In this perspective, we evaluate what automation should and can achieve for scientific practice. In doing so, we outline the current state of science automation, drawing on recent examples from different domains of science. Furthermore, we examine technological advancements that open new avenues for automation in science, and discuss current bottlenecks. Finally, we highlight a selection of practical and ethical considerations, and discuss how automation may lead scientists to work in a different world, one where traditional methodologies are redefined and new meta-paradigms for science emerge.

\section*{What are the bounds of automating scientific practice?}

Scientific practice can be defined as the set of methods and processes used by scientists to acquire knowledge about the natural world. Automation, in its broadest sense, refers to the use of technology to perform tasks with minimal human intervention. In the context of scientific practice, automation specifically denotes the use of technological tools and systems to carry out scientific tasks or processes traditionally performed by human scientists. 

The bounds of automation within scientific practice hinge on at least two questions: First, is there a desire and justification for automating a given scientific practice? This question touches upon \textit{goal-related bounds}---the alignment of automation with the overarching goals of science. Second, what factors characterizing a scientific practice influence the feasibility of automating that practice? This aspect focuses on the \textit{technological bounds}, assessing the practicality and potential constraints of applying automation in science.

\subsection*{Goal-related bounds: what automation should (not) achieve}

Science is driven by normative and epistemic goals. Here, we discuss arguments for and against automation serving these goals.

The normative goals of science involve ethical, moral, and societal values guiding both basic and applied science. One such goal may be to enable cheap and fast discoveries that advance human health. Along these lines, automation can serve to yield faster scientific discoveries with fewer resources. This is particularly desirable in the applied sciences, e.g., for identifying novel drugs or treatments. Thus, automation can aid scientific practice if societal needs are clear and research questions are well defined. However, the process of identifying a research question itself requires considering societal needs or the interests of the scientific community. As noted in the \textit{Opportunities} section below, generative artificial intelligence (AI) can integrate large bodies of literature to identify societally and scientifcally important gaps in our knowledge that are worth filling.
However, since the relevant normative considerations inherently depend on evolving human contexts, it can be argued that humans ought to always be involved in and monitor the degree to which scientific practices achieve these objectives \cite{binz_how_2023}.
Consequently, full automation in these areas might not only be impractical but also undesirable, underscoring the indispensable role of human scientists in addressing the normative dimensions of science.

The epistemic goal of science is to understand the natural world through description, prediction, explanation, and control. As discussed in the sections that follow, advances in machine learning can aid in automating the description or explanation of natural phenomena. Such automation can help reduce human errors and biases, leading to more accurate predictions and better control of natural phenomena. Even more so, automation may help bypass or augment the cognitive capacities of human researchers \cite{dubova_cognitive_2022},  enabling degrees of prediction and control unachievable for human cognition alone. For example, machine learning models can generate millions of proposals for novel materials that lie beyond human intuition  \cite{merchant_scaling_2023}. Yet, the increase in precision achieved through automation presents an epistemic dilemma, as automation can limit human understanding. In the basic sciences, advancement of human understanding may be more desirable than merely improving predictability through automation. The complexity of a machine learning model, for example, might enhance its ability to accurately predict new stable materials, but concurrently obscure the process by which these predictions are made for human scientists. This scenario illustrates a potential conflict between the scientific objectives of enhancing prediction, on the one hand, and enabling human understanding, on the other (see \textit{Practical Implications}). This suggests keeping human scientists involved in the scientific process rather than minimizing their involvement. Meanwhile, in applied sciences and engineering, the focus might shift towards maximizing prediction and control, providing a stronger case for automation of scientific practice.

\subsection*{Technological bounds: what automation can (not) achieve}

\begin{figure}
\centering
\includegraphics[width=.9\linewidth]{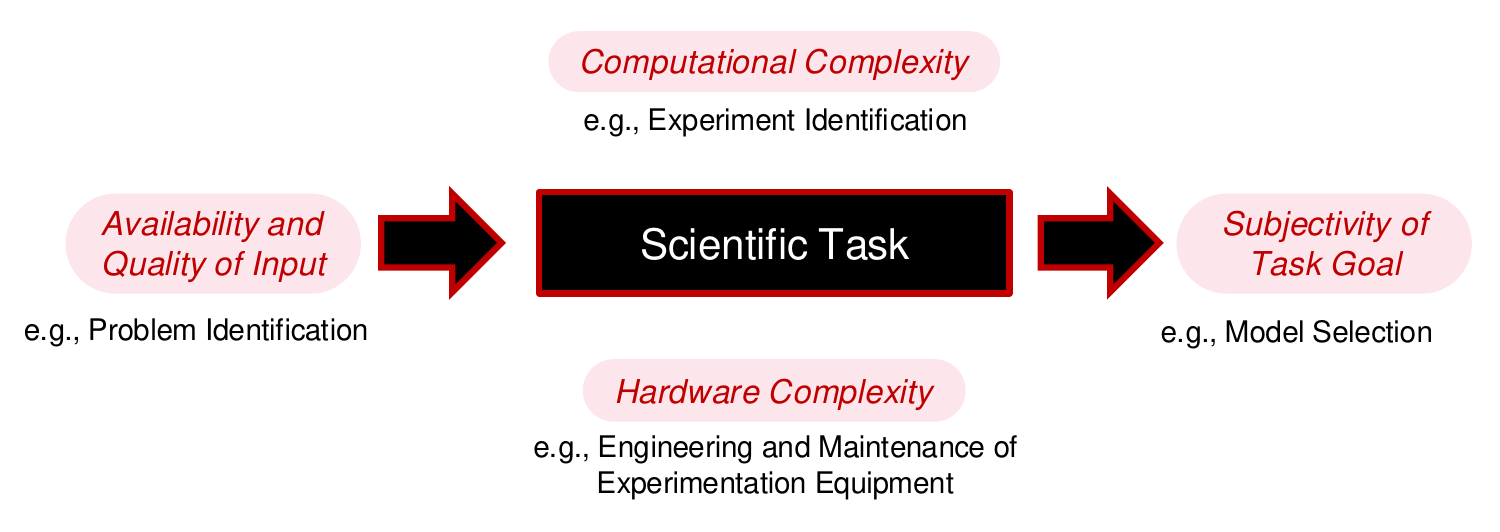}
\caption{Factors determining the technological reach of automation in scientific practice.}
\label{fig:factors}
\end{figure}

The technological bounds of automation hinge on the difficulty of automating scientific tasks. Here, we discuss four factors characterizing this difficulty (Figure \ref{fig:factors}). These factors indicate both opportunities and barriers to automation, thereby guiding the identification of areas within scientific practice where automation can be most effectively implemented or where it may face challenges. 

The first factor concerns the \textit{availability and quality of inputs} that a scientific task requires. Some tasks, such as identifying a research question, rely on diverse and sometimes subjective inputs, including peer opinions, news articles, or funding announcements. Such inputs may not be trustworthy, widely accessible or structured for machine processing, posing a challenge to automation.

Another limiting factor for automation is the \textit{computational complexity} of algorithms available to perform a scientific task. For example, identifying an appropriate experiment for testing a research question may require taking into account numerous decision variables (e.g., internal validity, resources needed, novelty) and searching an exponentially increasing space of possible experimental paradigms, which can be computationally intractable.

A related, yet often overlooked, factor influencing the automation of scientific tasks is the \textit{complexity of required hardware engineering}. As stated in Moravec's paradox, sensorimotor tasks, like executing invasive brain recordings or social experiments, require advanced solutions in robotics to facilitate automation, which can pose more significant challenges to automation compared to cognitive tasks \cite{moravec1988mind}. 

Finally, some tasks are difficult to automate because of the \textit{subjectivity of the task goal}. Some scientific goals cannot be easily turned into a well-defined objective, which is required to communicate it to a machine. For instance, choosing between scientific models can be a matter of personal preference \cite{dubova_ockhams_2024}. 

While the four factors collectively dictate the automatability of scientific tasks, they can be considered interdependent. For example, the automated discovery of scientific equations long relied on search methods with high computational complexity, such as evolutionary computation or brute force search, to identify a set of equations that best describes a given data set \cite{langley_scientific_1987,dzeroski_computational_2007}. However, the ability to collect large datasets cheaply, paired with improvements in computing hardware, enables the application of ``data-hungry'' but computationally tractable machine learning algorithms for equation discovery \cite{udrescu_ai_2020,cranmer_interpretable_2023,landajuela_discovering_2021,li_gfn-sr_2023}. This approach reduces computational complexity, illustrating how enhancements in one factor can compensate for limitations in another.

\section*{Automation in current scientific practice}

Existing approaches to automating science target tasks with readily available inputs, computational complexity and hardware demands that align well with current technological capabilities, and clear task goals. Accordingly, efforts at automatization in science have mostly been confined to tasks characterized by clearly specified objectives and well-defined subtasks, which include instances of quantitative hypothesis generation, experimental design, data collection, and quantitative analysis and inference. While covering all advances is out of the scope of this article, we highlight a subset of these approaches, focusing on cases that facilitated novel discoveries. 

\subsection*{Hypothesis generation}

Hypothesis generation is the development of testable statements that are based on observations, existing knowledge, or theory. Advances in automated hypothesis generation were primarily driven by two factors: improvements in computer algorithms, and the availability of large datasets. 

Initial automated hypothesis formation approaches relied on symbolic reasoning systems. For example, in organic chemistry, logical deduction based on existing knowledge was employed to formulate hypotheses about the chemical constituents of body fluids \cite{lindsay_dendral_1993}. Furthermore, quantum simulations, facilitated through cloud computing, became the backbone of hypothesis generation for materials properties \cite{saal_materials_2013,jain_commentary_2013}. The development of efficient search algorithms further expanded the scope of automated hypothesis formation to areas with large hypothesis spaces \cite{raayoni_generating_2021}. For instance, hypothesis generation in mathematics leveraged efficient machine learning algorithms to identify novel conjectures about fundamental constants \cite{raayoni_generating_2021}. Finally, deep learning enabled more breakthroughs in chemistry. A landmark achievement in this area is AlphaFold, which predicts 3D protein structures from amino acid sequences, facilitating the development of drugs \cite{jumper_highly_2021}. 

The availability of large data sets led to further advances in automated hypothesis formation. One example is the field of biomedicine, where large gene databases led to a surge in hypothesis generation with computational methods, e.g., using data mining and network analysis to propose genes that may be linked to diseases \cite{liekens_biograph_2011,voytek_automated_2012}. Similarly, existing materials databases provided sufficient information for machine learning methods to generate over 2.2 million proposals for novel materials that, so far, escaped human intuition \cite{merchant_scaling_2023}. 

\subsection*{Experimental design}

The problem of automated experimental design is to systematically identify the most informative experiment to address a particular hypothesis or scientific question. The informativeness of an experiment can be evaluated in various ways. Some automated experimental design methods are geared towards identifying the experimental conditions that minimize the influence of nuisance variables-–experimental variables that are not of interest but can pollute the informativeness of intended experimental manipulations \cite{musslick_sweetpea_2020,van_casteren_mix_2006}. Other methods aim to find experimental conditions that are well suited to identify a scientific model of interest \cite{navarro_assessing_2004,myung_optimal_2009,cavagnaro_adaptive_2010}. This problem of experimental design is closely related to the problem of active learning in machine learning research \cite{king_automation_2009,musslick_evaluation_2023,williams_cheaper_2015,coutant_closed-loop_2019}, which seeks to identify data points that can best inform a machine learning model when included as training data. A prominent active learning method used for scientific practice is Bayesian optimal experimental design, which has been successfully applied in various fields, including psychology \cite{cavagnaro_adaptive_2010,myung_optimal_2009,watson_quest_2017,valentin_designing_2023}, neuroscience \cite{shababo_bayesian_2013}, physics \cite{dushenko_sequential_2020,huan_simulation-based_2013}, biology \cite{kanda_robotic_2022,stanton_accelerating_2022}, chemistry \cite{korovina_chembo_2020,griffiths_constrained_2020}, materials science \cite{zhang_bayesian_2020,kusne2020fly,liang_benchmarking_2021}, and engineering \cite{papadimitriou_optimal_2004}. For example, in the domain of psychology, Bayesian optimal experimental design led to the discovery of novel models of how humans discount the future relative to the present \cite{chang_data-driven_2021}. 

While automated experimental design methods can facilitate efficient data collection and strong inferences, their efficacy can be compromised if the underlying assumptions are violated or if the scientific model is incorrectly specified \cite{grunwald_inconsistency_2017,rainforth_modern_2023,sloman_chapter_2022}. This limitation led to unexpected findings in simulation studies, where random sampling of experimental conditions outperformed automated theory-driven approaches to experimental design \cite{dubova_against_2022,musslick_evaluation_2023}, and where uniform sampling outperformed adaptive approaches in learning continuous relationships \cite{gelpi_sampling_2021}. 

Another limitation of current approaches to automated experimental design pertains to their scope, as they focus on navigating a pre-defined space of experimental manipulations. Exploring novel research directions, however, often involves identifying completely new experimental manipulations \cite{dubova_explore_2024}.

\subsection*{Data collection}

Data collection, often a time-consuming and costly aspect of empirical research, is a significant bottleneck in scientific discovery. Accordingly, automated tools for data collection emerged as some of the most impactful innovations in accelerating the pace of science. These tools span a wide range of applications and fields: fitness trackers revolutionized public health studies \cite{shin_wearable_2019}, continuous glucose monitors are providing critical insights into nutrition and diabetes research \cite{juvenile_diabetes_research_foundation_continuous_glucose_monitoring_study_group_effectiveness_2010}, and automated weather stations enhanced meteorological predictions \cite{lazzara_antarctic_2012}. In addition to providing streams of real-time data for ongoing analysis, these automated systems can minimize human observation and experimenter biases. Experimenter bias occurs when the beliefs, expectations, or preferences of the researcher unconsciously influence the conduct or outcome of an experiment. Automating data collection in animal studies helped to eliminate experimenter bias, resulting in refutations of previous results, such as the evidence for statistical learning ability in newborn chicks \cite{wood_automated_2019}. A particularly noteworthy advancement in the behavioral sciences was the adoption of web-based experiments, especially during the COVID-19 pandemic. Online platforms and interfaces for recruiting and conducting experiments did not only facilitate the collection of behavioral data at a time when traditional lab-based studies were impractical, but they also broadened the scope and diversity of participants \cite{gureckis_psiturk_2016,mason_conducting_2012,palan_prolific_2018}. Automating data collection also generated opportunities for automating other elements of behavioral science, such as adopting adaptive experimental designs that change based on the responses of participants \cite{thompson2022complex} or collecting larger datasets that can support the use of machine-learning algorithms \cite{peterson_using_2021}.

\subsection*{Statistical inference}

The automation of statistical inference transformed dramatically from the era of manual computations, a reality echoed in old statistical textbooks filled with computation-simplifying shortcuts. The introduction of computers altered statistical methodologies, sometimes even leading to their replacement by machine learning techniques. For example, modern statistical inference engines, like Stan, leverage techniques such as Markov Chain Monte Carlo (MCMC) for efficient sampling of model parameters \cite{carpenter_stan_2017}. Tools for likelihood-free inference enable the analysis of statistical models that are not mathematically tractable. Furthermore, frameworks such as Bayesian Workflow \cite{gelman_bayesian_2020} and platforms such as the Automatic Statistician \cite{steinruecken_automatic_2019} are streamlining complex processes like Bayesian inference and the construction of traditional statistical models. The automation of statistical inference, however, is mostly confined to the deduction of new knowledge based on pre-specified statistical models. 

\subsection*{Scientific inference and model discovery}

Scientific inference, unlike statistical inference, involves generating hypotheses about observations (abduction) and generalizing from observations to laws or broader theories (induction). The automation of scientific inference is termed computational scientific discovery and has so far centered on identifying models or laws that elucidate specific phenomena \cite{langley_scientific_1987,dzeroski_computational_2007,gobet_introduction_2019}. One instance of computational scientific discovery involves the identification of equations (``symbolic regression'') to uncover quantitative laws governing a given data set. Early efforts relied on heuristic search techniques to rediscover insights from mathematics \cite{falkenhainer_integrating_1986,lenat_ubiquity_1977} or physics \cite{langley_data-driven_1981}. Advances in machine learning and high-performance computing facilitated equation discovery, building on reinforcement learning \cite{landajuela_discovering_2021}, genetic algorithms \cite{cranmer_interpretable_2023,bartlett_genetic_2023,frias-martinez_automatic_2007}, MCMC sampling \cite{guimera_bayesian_2020}, mixed-integer nonlinear programming \cite{cornelio_combining_2023}, or gradient-based search techniques \cite{udrescu_ai_2020,li_gfn-sr_2023,landajuela_unified_2022,musslick_recovering_2021}. However, most forms of computational model discovery are limited to the rediscovery of existing knowledge. Possible exceptions include the discovery of scaling laws and boundary equations in plasma physics \cite{murari_data_2020} and novel models of human decision-making \cite{peterson_using_2021}. 

\subsection*{Closed-loop automation spanning multiple scientific practices}

\begin{figure}
\centering
\includegraphics[width=0.95\linewidth]{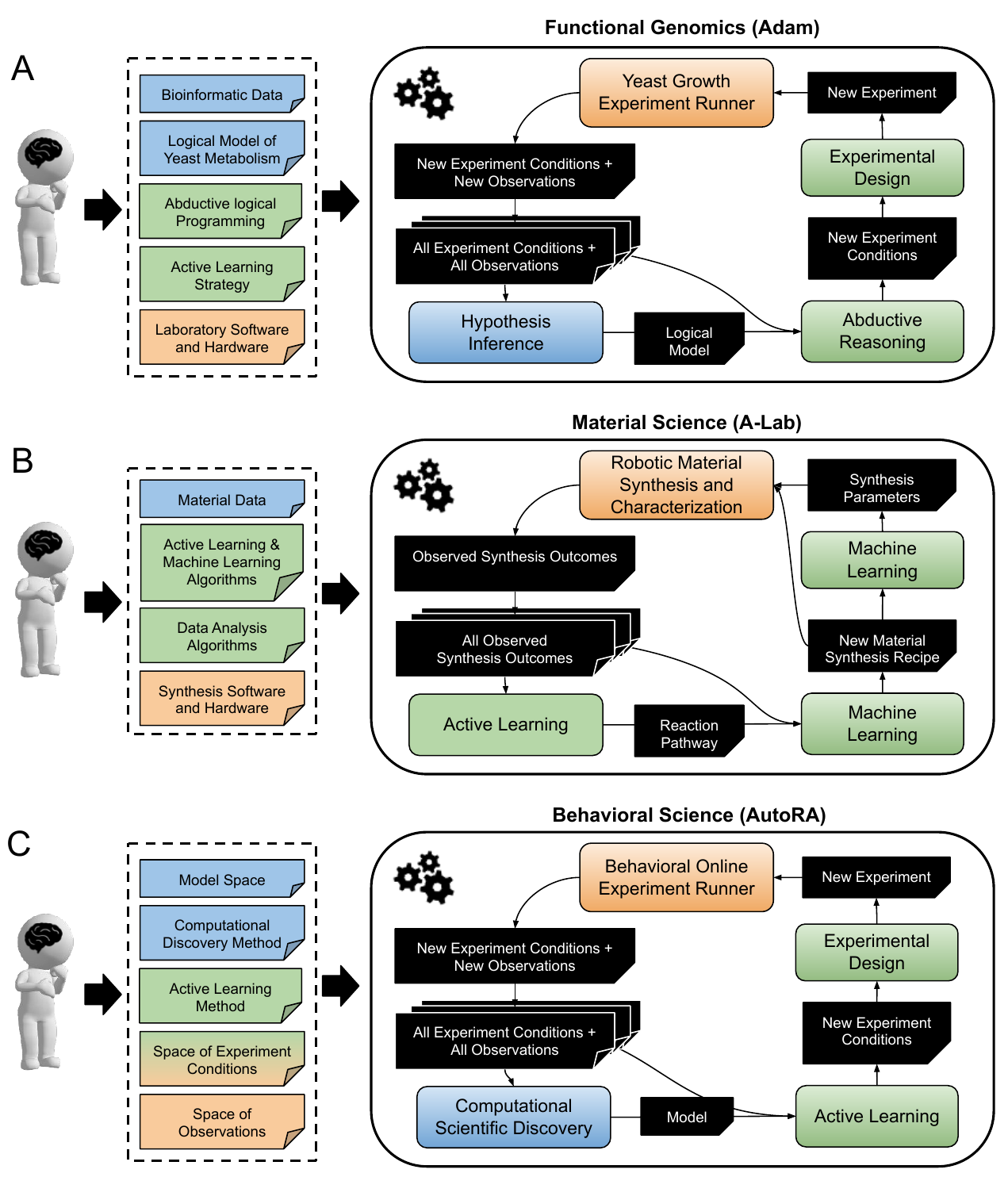}
\caption{Closed-loop automation systems. (A) Adam for functional genomics. (B) A-Lab for materials science. (C) AutoRA for behavioral science. Dashed boxes list knowledge and processes provided by human researchers.}
\label{fig:loop}
\end{figure}

Demonstrations of successful closed-loop automation in empirical research---implementing iterations between experimental design, data collection and model discovery---mark a significant progression for automated scientific practice. One pioneering example is the robot scientist Adam (Figure \ref{fig:loop}A), which was the first fully automated machine to discover novel scientific knowledge \cite{king_automation_2009}. Adam investigated the functional genomics of the yeast S. cerevisiae, and discovered the function of locally orphan enzymes---enzymes known to be in yeast but for which the gene(s) encoding them were unknown. The successor of Adam, Eve, is a robot scientist designed for early-stage drug development \cite{williams_cheaper_2015}, which identified chemical compounds that outperformed standard drug screening. Eve’s most significant discovery is that triclosan (an antimicrobial compound commonly used in toothpastes) may aid against malaria \cite{williams_cheaper_2015,bilsland_yeast-based_2013,bilsland_plasmodium_2018}. Another example of a closed-loop discovery system in biology is Wormbot-AI, a platform designed to autonomously conduct experiments on the longevity of worms, capable of testing thousands of interventions annually \cite{lee_million-molecule_2023,pitt_wormbot_2019}.

Complete automation also gained momentum in materials science and chemistry, where efforts are focused on integrating hypothesis generation, decentralized experimentation, and cloud-based decision-making. For instance, modular robotic platforms, driven by machine learning algorithms, were used to optimize material properties by varying synthesis conditions \cite{tamasi_machine_2022,macleod_self-driving_2020,pogue2023closed}. One notable example is A-Lab (Figure \ref{fig:loop}B), an autonomous laboratory for the solid-state synthesis of inorganic powders, which leverages a combination of active learning and machine learning models trained on the literature, to propose novel material candidates \cite{szymanski_autonomous_2023}. 

Additionally, behavioral research became amenable to closed-loop automation with the ability to collect data via online experiments. Open-source tools like AutoRA \cite{Musslick_AutoRA_Automated_Research_2023} facilitate closed-loop research by integrating automated model discovery, experimental design, and experimentation in empirical research. AutoRA effectively interfaces with web-based platforms for automated data collection, integrating the acquisition of behavioral data from human participants. While the potential to yield novel discoveries stands to test, AutoRA served as a computational testbed for philosophy of science, exposing cases where random experimentation outperforms model-guided experimentation \cite{musslick_evaluation_2023}.

Finally, researchers introduced an LLM-based agent for automating empirical machine learning research, from idea development and experimental design to execution and data analysis, e.g., for improving existing machine learning models \cite{lu2024aiscientist}. Notably, this system also leveraged LLMs to automate the writing and peer review of the resulting research manuscript, with the computational cost of one article estimated to be just 15 USD.

Despite their potential to accelerate scientific discovery, it is important to recognize that the pioneering examples of closed-loop automation are currently confined to specific, automatable research steps and operate within a constrained range of experimental design and model spaces as delineated by human researchers.

\section*{Future opportunities} 

Existing approaches for automating scientific practice primarily target tasks for which (a) high-quality data is available, (b) the computational complexity can be addressed by current algorithms, and (c) hardware complexity is manageable. The most promising prospects for future automation in scientific practice are found in tasks traditionally limited by human cognitive capacities. This includes areas requiring the processing of large volumes of high-dimensional data or exhaustive literature searches. In this section, we highlight a few technological trends that promise to push the boundaries of science automation along these lines. 

\subsection*{Data collection, standardization, and sharing}

Advancements in cost-effective data collection, standardization, and sharing significantly boost the automatability of scientific practices, particularly those dependent on empirical data. For example, in the behavioral sciences, the utilization of crowd-sourced experimentation platforms like Amazon Mechanical Turk and Prolific revolutionized the efficiency of behavioral data collection. Additionally, LLMs that can mimic human behavior were proposed as proxies for participants, aiding in the acquisition of large-scale datasets \cite{dillion2023can}.
Once acquired, such large---yet cost-efficient---datasets can empower data-hungry machine learning algorithms, enabling them to uncover novel, and more precise models of human behavior \cite{petersen2021deep,almaatouq_beyond_2022,eckstein_predictive_2023,ji-an_automatic_2023}. Large-scale data collection, however, still bears significant hardware challenges, e.g., for collecting biological samples from a large number of participants (see \textit{Future challenges}). Nevertheless, the data quality needed for automated analysis techniques should be complemented by data standardization and sharing. 

Scientific data sharing platforms, such as the Open Science Framework, facilitated the availability and accessibility of data needed for automated analyses and computational discovery. The potential of data sharing and standardization is perhaps best illustrated in materials science, where databases for stable materials enabled the prediction of large quantities of new materials \cite{merchant_scaling_2023}. Other scientific domains profit from similar efforts. For example, in neuroscience, archives like DANDI, OpenNeuro, DABI and BossDB allow researchers to share data using community standards \cite{subash2023comparison}, such as  BIDS for neural data \cite{gorgolewski_brain_2016}. 


\subsection*{Combining data-driven and knowledge-driven discovery}

A particularly promising approach to automating scientific discovery is the integration of pre-existing human knowledge into the discovery process. Traditionally, data-driven discovery methods operated with minimal prior knowledge about the specific domain of scientific inquiry. This pure data-driven approach makes such methods particularly susceptible to noisy data. However, recent work demonstrates that incorporating prior theoretical knowledge can significantly aid in recovering scientific models from noisy datasets. For example, Bayesian symbolic regression exhibits greater efficacy in recovering equations from noisy data when given priors about scientific equations extracted from Wikipedia \cite{guimera_bayesian_2020,hewson_bayesian_2023}. Similarly, embedding prior knowledge in the form of general logical axioms proved instrumental in rediscovering complex scientific laws, including Kepler’s third law of planetary motion and Einstein’s relativistic time-dilation law \cite{cornelio_combining_2023,cory2024evolving}. Furthermore, experiments with the BacterAI, which uses active learning for the automated study of microbial metabolisms, have demonstrated the advantage of leveraging relevant prior knowledge \cite{dama2023bacterai}. Specifically, when the metabolic model trained on one bacterial species was retrained for the species of interest, it more efficiently discovered its metabolic model compared to starting the learning process from scratch, despite the two species differing in their metabolic capabilities. These examples highlight the benefits of combining data-driven and knowledge-driven approaches for automated model discovery.

The benefits of knowledge-driven discovery are, however, fundamentally limited by the quality of prior knowledge. For example, Bayesian adaptive experimentation can be misled if prior knowledge mischaracterizes the data \cite{sloman_characterizing_2022,sloman2024knowing}. Thus, data-driven approaches to computational model discovery become particularly beneficial when dominant scientific models in the empirical sciences are more informed by (wrong) theory versus data. This is evident in computational models of human reinforcement learning, which predominantly rely on classic machine learning algorithms \cite{sutton_reinforcement_2018}. Recent work demonstrated that a data-driven model discovery can uncover novel reinforcement learning models that better explain human learning than traditional models \cite{eckstein_predictive_2023}. 

Finally, a notable area of progress in automated model discovery is the analysis of high-dimensional datasets, such as fluid dynamics captured in video format, through reduced-order modeling. This process involves learning a low-dimensional representation of the dynamics inherent in complex data and then decoding the governing equations of these latent dynamics \cite{kutz_machine_2023,conti_reduced_2023,mendible_dimensionality_2020,chen_automated_2022}. 
Similar approaches were developed to automate the discovery of neural data embeddings correlating with behavioral dynamics \cite{schneider_learnable_2023}. 
These approaches promise to extend the reach of automated model discovery to high-dimensional naturalistic datasets. beyond experimental control.

\subsection*{Generative AI and LLMs}

Generative AI and LLMs offer paths towards automating scientific practices that have historically been challenging due to their computational complexity and qualitative nature \cite{boiko_autonomous_2023,birhane_science_2023,lu2024aiscientist,zheng_large_2023}. Among these are the synthesis and integration of literature, and documentation of findings. 

Researchers argued that LLMs show promise in enhancing literature reviews, a task currently limited by the cognitive constraints and language barriers of human scientists \cite{guler_artificial_2023,whitfield_elicit_2023}. Whereas humans may only be able to parse and integrate a few hundred articles into a literature review---the scope of which is heavily influenced by the expertise and biases of the researcher---LLMs may accomplish literature synthesis in the order of thousands or millions of articles. Critically, LLMs can take into account articles written in different languages, thus helping to counter the dominance of Western perspectives in scientific literature. Thus, LLMs can assist in extending or even bypassing  human researchers’ cognitive limitations. A notable application of LLMs for the purpose of literature synthesis is Elicit, which utilizes LLMs trained on paper abstracts to support and help researchers extract relevant information from the scientific literature \cite{whitfield_elicit_2023}. Another instance of such assistance is an LLM-based ``co-scientist'' for chemical research, which improved the planning of chemical syntheses based on extensive information available on the internet, and aided in the navigation of extensive hardware documentation \cite{boiko_autonomous_2023}. Additionally, BrainGPT---an LLM fine-tuned to the neuroscience literature--- demonstrated the capability to outperform human experts in predicting the results of neuroscience experiments \cite{luo2024large}.

Combined with their capability for literature synthesis, LLMs can foster the discovery of new research directions and hypotheses \cite{lu2024aiscientist}. Along these lines, LLMs have the potential to expand experimental design spaces, addressing a common bottleneck in automated scientific practice. While traditional automated experimentation is confined to researcher-defined variables (cf. Figure \ref{fig:loop}), LLMs could identify novel experimental variables of interest, thus broadening the scope of scientific inquiry. However, it can be argued that LLMs risk rediscovering already known hypotheses and experiments \cite{binz_how_2023}. 

Once experiments are designed, LLMs may aid in the balanced documentation and communication of the research study, including the automated documentation of research code \cite{chen_evaluating_2021,su_hotgpt_2023}. Apart from aiding in the construction of research articles, LLMs can enable automated translation into multiple languages. This advancement is particularly beneficial for non-native English speakers and is an example of how automation and AI can address ethical challenges in science. Nevertheless, literature reviews conducted by human scientists serve not only to synthesize knowledge but also to build and refine the conceptual frameworks of evolving scientists---a process that is critical to scientific training and that is challenged by the overuse of LLMs for literature synthesis.

\section*{Future challenges}

Despite recent advances and opportunities for the automation of science, there remain substantial obstacles. This section examines technological bounds rooted in four bottlenecks (cf. Figure \ref{fig:factors}): limited availability and quality of data, intractable computational complexity of certain scientific tasks, lack of required hardware, and subjectivity in assessing the outputs of scientific tasks. These bottlenecks highlight why barriers to automation remain difficult to surmount in the basic sciences (as opposed to engineering), at least with the technologies and methodologies currently at our disposal. Addressing these challenges will require significant interdisciplinary efforts to identify solutions that enable automation beyond a few selected domains of scientific inquiry.

\subsection*{Limited availability and quality of inputs}

Prior applications of computational discovery, such as in chemistry \cite{de_almeida_synthetic_2019,lindsay_applications_1980,henrich_weirdest_2010} and materials science \cite{merchant_scaling_2023,szymanski_autonomous_2023}, relied on standardized formats for both data and scientific hypotheses that are easily parsed by machine learning algorithms. However, most tasks of scientific practice rely on a diversity of representations for scientific knowledge. For example, computational models in the natural sciences are expressed in various formats, such as equations embedded in scientific articles or computer code written in different programming languages. Without standardization across disciplines, automated systems face significant challenges in drawing parallels or applying concepts from one domain to another. Efforts to standardize the representation of scientific models and other forms of scientific knowledge promise to ease the automation of scientific practices relying on such knowledge \cite{gleeson_integrating_2023}. However, even if data is standardized and widely available, ensuring its quality remains critical. For instance, literature synthesis enabled by LLMs may be unfruitful or even misleading if fraudulent or unreproducible papers are included as inputs to these models. Therefore, robust quality control measures must accompany standardization efforts to maintain the integrity and usefulness of automated systems. 

\subsection*{Computational complexity}

One of the fundamental bottlenecks in the automation of scientific practice lies in the computational complexity of many scientific tasks. For example, complexity analyses within the realm of cognitive science indicate that scientific discovery in cognitive science may be computationally intractable in principle, even with unlimited availability of data \cite{rich_how_2021}. These theoretical results suggest that uncovering a definitive ``ground-truth'' theory may be beyond the reach of computation. 

One potential critique of leveraging computational methods for scientific discovery hinges on the incomplete comprehension of the cognitive processes, and the concomitant computational complexity underlying it. One may argue that without a full grasp of how humans tackle scientific inquiries, designing algorithms capable of similar feats seems implausible. However, at least two counterarguments challenge this perspective. First, replicating natural processes is not a prerequisite for solving problems. For instance, modern airplanes achieve superior lift not by emulating the flapping motion of birds but through aerodynamically efficient designs. Second, a deep understanding of cognitive phenomena is not a strict requirement for automation, as evidenced by the capabilities of LLMs to produce coherent natural language sequences without humans having a complete scientific understanding of language generation. Nonetheless, this gap in understanding underscores the importance of implementing robust evaluation methods to ensure the accuracy and mitigate any potential negative impacts of automating scientific processes.

\subsection*{Hardware engineering}

The advancement of automated science is significantly hindered by current limitations in laboratory robotics and hardware engineering. For instance, executing complex biological or physics experiments remains challenging. Moreover, while robotic automation has been successfully implemented in certain areas, such as with the robot scientist concept \cite{king_functional_2004,king_automation_2009,burger_mobile_2020,dama2023bacterai}, its application is primarily limited to clearly defined engineering problems. Yet, even well-defined engineering problems must manage the noise and variability inherent in the data collected by sensors, which can dramatically affect the reliability of scientific outcomes.
Therefore, while progress has been made in automating scientific practice, developing more sophisticated robotics to handle complex, noisy data is crucial for its broader adoption and effectiveness.

The automation of hardware tasks in scientific practice is also hindered by the need for highly specialized equipment, leading to significant capital expenditures, often exceeding millions of dollars. Such custom-built hardware is typically field-specific and lacks versatility for reuse in other scientific domains. This challenge is evident in the limited cross-utilization of hardware between disciplines, as seen in the relatively small amount of equipment that materials scientists have been able to adapt from the more heavily automated field of drug discovery. Addressing this issue requires a strategic approach where, for each scientific field, scientists identify and develop a core set of automated hardware that can deliver the greatest impact. This not only involves designing equipment that meets the unique needs of each field but also balancing specificity with adaptability, to maximize utility and cost-effectiveness.

\subsection*{Subjective goals of scientific tasks}

More than in engineering, practices in basic science are inherently subjective in how the outcomes of those practices are evaluated. This challenge is particularly evident in developing AI capable of generating novel and impactful scientific ideas. Novelty and impact involve a high degree of subjectivity and variability, making it difficult for these systems to replicate human judgment in the space of scientific inquiry \cite{birhane_science_2023}. This issue is compounded by the personal aspect of scientific practice. The selection of scientific projects is guided by the personal experience and perspective of human scientists. Diversity in such perspectives paired with interdisciplinary exchange can lead to a greater diversity of ideas in human scientific systems \cite{messeri2024artificial}---a dimension that AI currently cannot emulate without explicit instruction. Furthermore, the lack of standardized solutions in many scientific areas means that automating these tasks risks constraining exploration, which is vital for scientific advancement. 

Moreover, interpretation of data patterns and hypothesis generation often necessitates human judgment to translate statistical regularities into meaningful scientific interpretations. Techniques like topic modeling, while effective in identifying text co-occurrence patterns, require human insight to align these patterns with relevant scientific constructs \cite{chang_reading_2009}. 
The role of human judgment is perhaps best exemplified in serendipitous discovery, often stemming from unexpected failures or results. For example, Alexander Fleming's discovery of penicillin began with the accidental contamination of a Petri dish. Instead of discarding it, his observation of the bacteria being killed by the mold led to the development of the first antibiotic. These aspects highlight the crucial role of human judgment in scientific discovery.

\section*{Implications}

Although the automation of science currently faces significant limitations, the extent to which it will evolve in the mid- to long-term remains an open empirical question. As advancements in hardware and algorithms continue, the range of practices subject to automation is likely to expand. In this section, we explore the practical and ethical consequences of this trend.

\subsection*{Practical implications}

\subsubsection*{The role of human scientists and the paradox of automation}

The advancement of automation in scientific practice raises considerations regarding the future role of human scientists. On the one hand, it can be argued that automation reduces the need for human involvement. Scientific discovery systems may become able to monitor themselves and tune themselves to optimal performance---potentially excluding humans from the scientific discovery loop. On the other hand, it can argued that the greater the efficiency of an automated system, the more vital the role of human oversight \cite{bainbridge_ironies_1983}. A critical assumption underlying this ``paradox of automation'' is that automation is not perfect; the potential for accumulating errors necessitates human intervention. If automation were flawless, human oversight would be unnecessary, and the paradox would not exist. However, for tasks with sufficient complexity and uncertainty, this paradox suggests that, in highly automated environments, human contributions, though less frequent, are more critical. This may specifically apply to tasks that demand subjective assessment or the synthesis of complex data, such as reviewing scientific literature, as well as high-level responsibilities such as strategic allocation of funds for scientific inquiry. 

Even in the absence of subjective assessment, there are inherent risks associated with automation. For instance, an error within an automated system can lead to a cascade of compounded errors, persisting and potentially amplifying until the system is either corrected or deactivated. This may be particularly problematic for automation methods whose decision-making processes are not completely predictable, as is the case for many machine learning algorithms. This unpredictability raises the issue of responsibility for unintended consequences such as injuries. Given the potential severe legal and financial implications of compounding errors in automation, the involvement of human scientists, even in areas where automation is technically feasible, may prove to be more efficient, practical, and safe in the near future. Thus, the paradox of automation underscores the lasting importance of human expertise and the need for a balanced approach that combines automated systems with human judgment.

\subsubsection*{Research training}

With increased automation of science, there arises a need to reevaluate and adapt scientific education. This new landscape calls for training that encompasses not only traditional scientific knowledge but also skills for effectively working alongside automated scientific discovery systems. For instance, obtaining valuable outputs from LLMs is becoming an essential skill. Moreover, scientists will need to develop competencies in understanding and evaluating the functioning and outputs of automated systems, as is already demanded for statistical software \cite{stanton_accelerating_2022}.  This shift implies a growing demand for engineers, scientists, and technicians proficient in advanced STEM skills.

\subsubsection*{Research evaluation}

The current pace of science is primarily determined by our capacity to carry out the research itself. Laboratory studies in fields like biology and chemistry can take years, contrasting with the relatively quick peer review process. However, if advancements in automation enable research to be conducted and documented several magnitudes faster \cite{lu2024aiscientist}, this could lead to a substantial increase in the rate of research article submissions. Such a scenario would further strain the already pressured peer review system. One potential solution could be the automation of peer review, possibly through the use of LLMs; however, this approach has already faced restrictions and bans in certain contexts due to concerns about its efficacy, reliability, and confidentiality \cite{national_institutes_of_health_use_2023}. Another potential solution is for journals to require that articles generated by automated systems be accompanied by critical evaluations from corresponding human authors. This ensures that human researchers retain comprehension and oversight of what is being submitted while also serving as initial reviewers of the work generated by their automated systems.
Either way, this shift would necessitate a reevaluation of the peer review process, ensuring it remains rigorous and effective in the face of increased scientific productivity. 

\subsubsection*{Scientific methods}

The automation of scientific practice has the potential to bring about a shift in scientific methods that goes beyond mere acceleration of scientific discovery. As discussed above, the use of machines for scientific discovery allows us to move beyond the cognitive and physical constraints inherent to human scientists \cite{dubova_cognitive_2022}. 
Consider, for example, the principle of parsimony in the construction of scientific models.
Traditionally, parsimonious models have been favored for their superior generalization, ease of interpretation and communicability among human scientists.
However, as discussed in \citep{dubova_ockhams_2024}, recent studies suggest that highly complex models can, under certain conditions, surpass the generalization capabilities of simpler ones \cite{belkin2019reconciling}, leading to unprecedented advances in scientific research (e.g., for 3D protein folding \cite{jumper_highly_2021} or material discovery \cite{merchant_scaling_2023}). Moreover, as explored in \citep{dubova_ockhams_2024}, the development of such complex models is often a prerequisite for discovering successful parsimonious models (e.g., \cite{frankle2018lottery,li2020train,agrawal_scaling_2020}). This ability of machines to explore and develop models with a level of complexity beyond what is readily interpretable by humans opens up new avenues for scientific progress, less constrained by human cognitive limitations. However, as discussed above, for basic science, there is epistemic value in \textit{human} understanding that may outweigh the predictive power of AI scientists.

Another consequence of automation concerns the ways in which empirical research is conducted. For example, automated systems can hypothesize and experiment in design spaces far beyond the reach of human cognitive capabilities \cite{merchant_scaling_2023,burger_mobile_2020}. Furthermore, the ability to collect large amounts of data cheaply may obviate frequent iterations between hypothesis generation, experimental design, and data collection. Instead, with the availability of large data sets, the problem of scientific discovery can be transformed into a model discovery problem more amenable to machine learning \cite{peterson_using_2021,griffiths_manifesto_2015,almaatouq_beyond_2022}. However, it is important to recognize that the success of a one-time large-scale data collection hinges on a well-defined experimental design space and the stability of the system under study, as constant changes in the system can undermine the effectiveness of this approach. Accordingly, adaptive experimental design may be needed to identify suitable design spaces \cite{dubova_against_2022}.

\subsection*{Ethical implications}

\subsubsection*{Biases}

While human biases influence every aspect of scientific work, automated systems are not immune to bias. They can inherit biases from their creators, the construction process, the data they use, and their training format \cite{daston_objectivity_2021}. Examples include discriminatory biases in facial recognition technology \cite{buolamwini_gender_2018}, unrepresentative sampling in psychological experiments \cite{henrich_weirdest_2010}, and discrimination in automated participant recruitment processes \cite{yarger_algorithmic_2020}. Moreover, automated literature reviews don't escape the biases inherent to the existing literature. These biases can be democratized and exacerbated by the pace of these systems, especially when they are uninterpretable or operate as ``black boxes.'' However, a potential advantage is that biases in automated systems may be easier to correct than in humans, such as by using more diverse data, or by aligning automated systems with societal norms.

\subsubsection*{Value alignment and responsibility}

The risk of harmful biases and outcomes of automated processes call for their value alignment with broader societal norms. This is particularly crucial as automation could potentially ease the path for malevolent entities to conduct research detrimental to society, such as developing chemical or biological weapons. Such outcomes underscore the necessity of ethics dedicated to addressing these issues, ensuring that automated scientific advancements align with human values. 

Consequences of automation also bring about the issue of responsibility: If a scientific discovery that affects the wider society is based on an automated process, who is responsible? The accountability for effects arising from harmful scientific practice remains ambiguous—whether it lies with the system's creator, its user, or the implementer of societal changes based on the system’s output. This issue parallels broader debates in AI, such as liability in self-driving car accidents or the creation of automated artwork. Additionally, the potential misuse of powerful systems (e.g., a system suggesting harmful drug treatments) necessitates robust safeguards. The same applies to potential violations of data privacy. When automated systems generate contentious theories or design ethically questionable experiments, human oversight and responsibility are imperative. Importantly, ethical guidelines are often formulated by the institutions developing the systems \cite{hagendorff_ethics_2020}, highlighting the need for an external framework that can hold institutions accountable.

\section*{Conclusion}

While the automation of scientific practice is currently confined mostly to well-defined engineering and discovery problems, there is the potential for automation to pervade a large part of scientific practice. We suggest that this trend represents not merely a series of quantitative changes, such as increased efficiency or precision in science, but brings about a fundamental shift in the conduct of science. The integration of AI into scientific practice has the potential to overcome human cognitive limitations, thereby expanding our capabilities for discovery. Yet, this advance is not without challenges---data availability, computational complexity, engineering demands, and subjectivity of scientific task goals mark the technical boundaries of current automatability. Furthermore, normative goals of science---anchored on societal values---potentially make complete automation of scientific practice neither desirable nor feasible. Finally, this qualitative shift comes with practical and ethical challenges that call for interdisciplinary and collective efforts from researchers, policymakers, and the broader community to navigate the future of science.

\section*{Disclosures}
The authors have no competing interests to report.

\acknow{S. Musslick and S. Mahesh were supported by Schmidt Science Fellows, in partnership with the Rhodes Trust. S. Musslick was also supported by the Carney BRAINSTORM program at Brown University and the National Science Foundation (2318549). S. Mahesh also acknowledges the support of the Acceleration Consortium fellowship. S.J. Sloman acknowledges support from the UKRI Turing AI World-Leading Researcher Fellowship, [EP/W002973/1]. S. Chandramouli was supported by the Finnish Center for Artificial Intelligence, and Academy of Finland (328813); he also acknowledges the support from the Jorma Ollila Mobility Grant by Nokia Foundation. L. Bartlett and F. Gobet were supported by European Research Council Grant ERC-ADG-835002-GEMS. T. L. Griffiths was supported by a grant from the NOMIS Foundation. R. D. King was supported by the Wallenberg AI, Autonomous Systems and Software Program (WASP) funded by the Knut and Alice Wallenberg Foundation, by Chalmers Artificial Intelligence Research Centre (CHAIR), and by the UK EPSRC grants EP/R022925/2 and EP/W004801/1. The authors thank Solomon Oyakhire for valuable feedback.}

\showacknow 

\bibliography{main}

\end{document}